# SRF Cavity Emulator for PIP-II LLRF Lab and Field Testing

Ahmed Syed, Brian Chase, Philip Varghese, Sana Begum,
RF Department,
Fermilab
Batavia, IL, USA

*Abstract*— **There are many stages in the LLRF and RF system development process for any new accelerator that can take advantage of hardware emulation of the high-power RF system and RF cavities. LLRF development, bench testing, control system development and testing of installed systems must happen well before SRF cavities are available for test. The PIP-II Linac has three frequencies of SRF cavities, 162.5 MHz, 325 MHz and 650 MHz and a simple analog emulator design has been chosen that can meet the cavity bandwidth requirements, provide tuning errors to emulate Lorentz force detuning and microphonics for all cavity types. This emulator design utilizes a quartz crystal with a bandwidth of 65Hz at an IF of ~ 4 MHz, providing a Q of ~ 1.3 x 10^7 at 650MHz. This paper will discuss the design and test results of this emulator.**

## I. INTRODUCTION

Low-level radio frequency (LLRF) systems are needed in accelerators to control the rf fields in the rf cavities that are used for beam acceleration. Generally, it is required to build and test these LLRF systems before commissioning the beam. Functional testing is frequently required for the study of various LLRF control algorithms during beam commissioning. However, the prerequisites for LLRF system testing, such as super-conducting surroundings and the appropriate rf sources, are typically pricey and occasionally unavailable.

State-space models of RF systems that are software-based are prevalent and have useful applications in the design phase of an LLRF system, but it is challenging to operate the full rf system in real time [5]. This issue led to the creation of cavity emulators to allow for LLRF system to be developed without the need to connect to a real superconducting rf cavity. Real-time behavior under LLRF control system operation can be achieved using this emulator. However, the research is largely concentrated on the cavity's fundamental mode [6]. Moreover, the performance of the LLRF system is limited by disturbances like microphonics and detuning which are in scope for this design.[1]

## II. LLRF SYSTEM

As shown in Fig. 1, rf devices, such as rf sources (such as klystron) and rf cavities, are essential in order to test the LLRF system. Real-time emulators must be available as a stand-in for actual cavities and RF sources in the development phase since these devices are not always readily available. The RF system, however, is prone to disturbances. Beam loading, high voltage power supply ripples, Lorentz force detuning (LFD), and unknown disturbances such microphonics, master oscillator phase noise, and clock jitters are examples of typical sources [3-4]. The LLRF system's performance is constrained by these disturbances, consequently, useful information can be obtained through the incorporation of disturbances into the cavity emulator. According to the above discussion, this paper discusses the design and implementation of a LLRF analog cavity emulator based that integrates both RF cavity models and RF disturbance models.

## III. CAVITY EMULATOR DESIGN

Design of cavity emulator involves optimization of crystal filter circuit in terms of achieving impedance matching and frequency tuning. Crystal filter circuit design is developed in ADS simulation. Once the crystal filter is optimized and developed frequency upconversion and downconversion is developed for IF – RF conversion.

Based on the input signals (RF drive) the cavity emulator generates the following output signals:

- Cavity Forward
- Cavity Reverse
- Cavity Probe

The basic idea of the emulator is to use a simple crystal filter at an IF 4.1938470 MHz and narrow bandwidth of ~77Hz. To achieve the highest Q possible, crystal filter like quartz is used, hence the Q factor of superconducting RF cavities that are used in accelerator field can reach as high as $10^{11}$. For the crystal board of 4.19 MHz, Q= 8.38 x $10^4$. Best effort is made to pick the crystal with least possible bandwidth as that would affect quality factor of overall response.

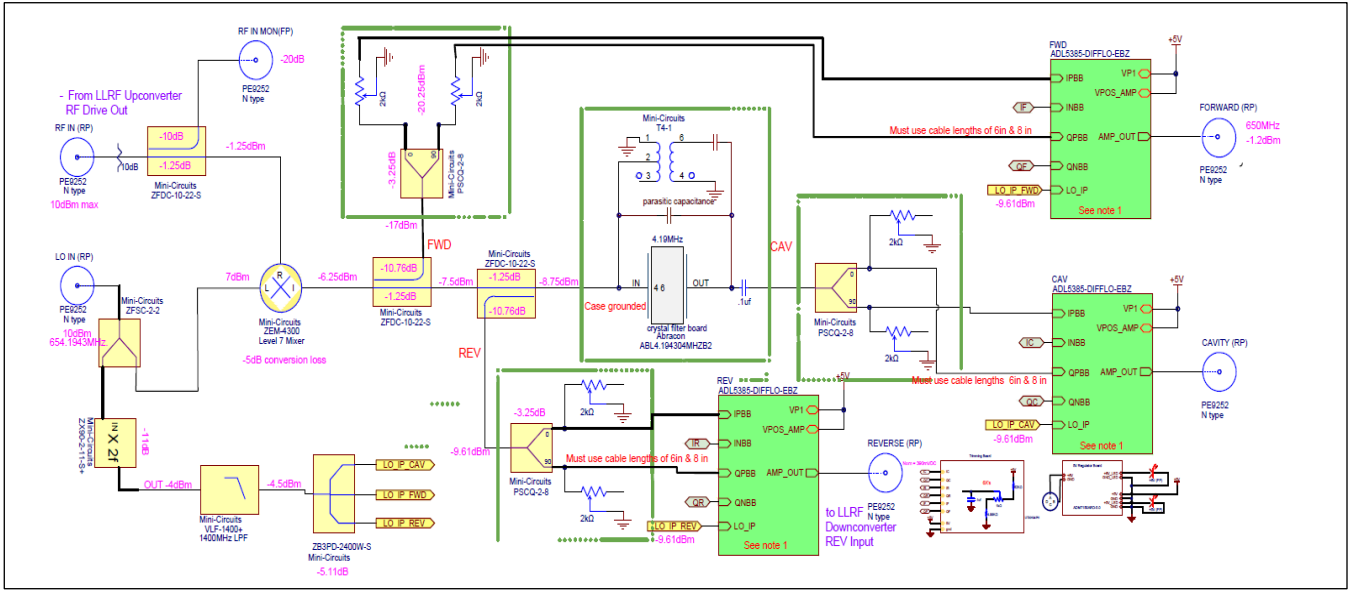

Fig.1 Detailed Schematic of Analog Cavity Emulator

*Design of the crystal board*

The crystal board forms the crux of the cavity emulator circuit and the output characteristics of the signal like- bandwidth and return loss from this board determines the final output characteristics. The crystal board is design involves DC blocking capacitors and coupling capacitors as shown in Fig.2. Firstly, RLC equivalent circuit of the crystal is determined by measuring the crystals series and parallel resonance frequency with a network analyzer. Upon obtaining the equivalent -Cs, Ls, Rs and Cp we incorporate this in the circuit, and this forms the quartz crystal equivalent.These obtained values of quartz filter equivalent circuit are used in the ADS model.

*Use of transformer for negative capacitance*

RF transformer T4-1 inverts signals by producing a 180-degree phase shift. It is in series with a equivalent capacitance equal to parallel capacitance of the crystal. This combination of RF transformer and series capacitor effectively cancels out the effect of parallel resonance peak which is undesired. This is done to improve the signal floor of the crystal side band and compensates for parasitic capacitance.

Fig. 4 below shows ADS simulation without RF transformer shows the parallel resonance peak $f_p$ is dominant and the series resonance peak $f_p$ has a lower S21 magnitude at the center frequency.

It also shows poor return loss at center frequency and doesn't depict the behavior of an RF superconducting cavity quite well.

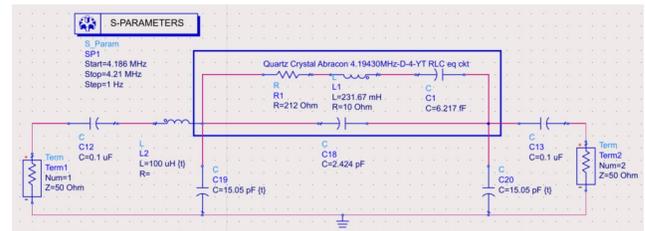

Fig. 2 ADS Circuit Schematic- without transformer

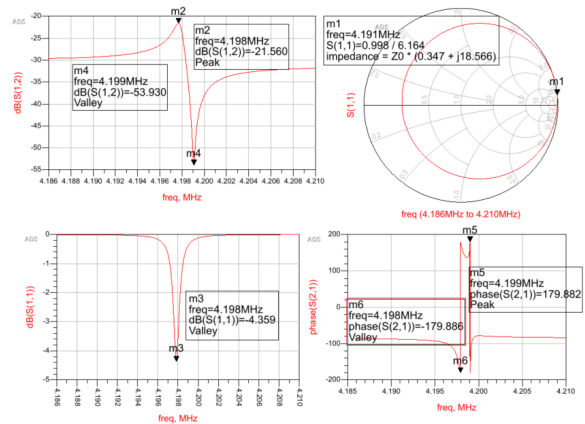

Fig. 3 ADS simulation plots - without RF transformer.

In order to eliminate the parallel resonance peak $f_p$ of the circuit we introduce RF transformer T4-1 which has center tapped primary winding. Signal at the input is inverted by 180° and is in series with capacitor C13 with almost same capacitance as parallel equivalent capacitance C18 of crystal filter. The overall response of the transformer and capacitor C13 effectively cancels out the effect of parallel equivalent capacitance of quartz crystal which is also known as case capacitance or parasitic capacitance.

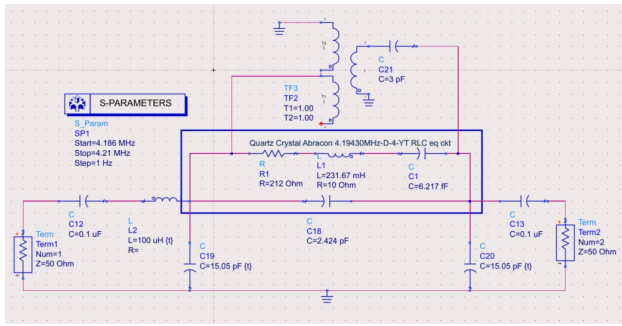

Fig. 4 ADS simulation plots -with RF transformer.

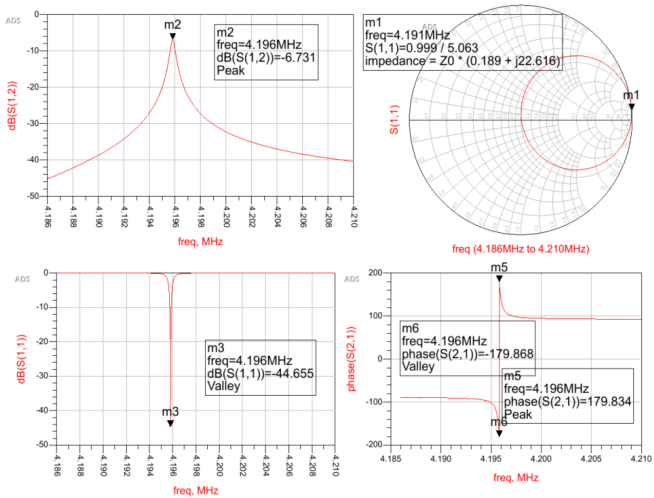

Fig. 5 ADS simulation plots – with RF transformer.

## Lorentz Force Detuning and Microphonics

The cavity emulator design demonstrates the capability to simulate Lorentz force detuning and microphonics. Crystal filter circuit is simulated in ADS, by varying input and output capacitance of the crystal filter center frequency can be moved. Varactor diodes are used to be able to vary the capacitance by applying external bias voltage. Upon including in the circuit varactor diodes would make the center frequency dependent on capacitance thereby external bias DC voltage. These DC signals can be provided with piezo control signals provided to the real superconducting RF cavity to tune the cavity off and on resonance.

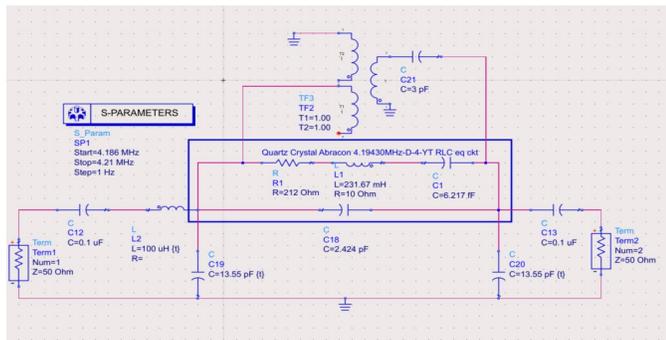

Fig. 6 ADS simulation schematic -with frequency detuned to 4.197MHz from 4.196MHz

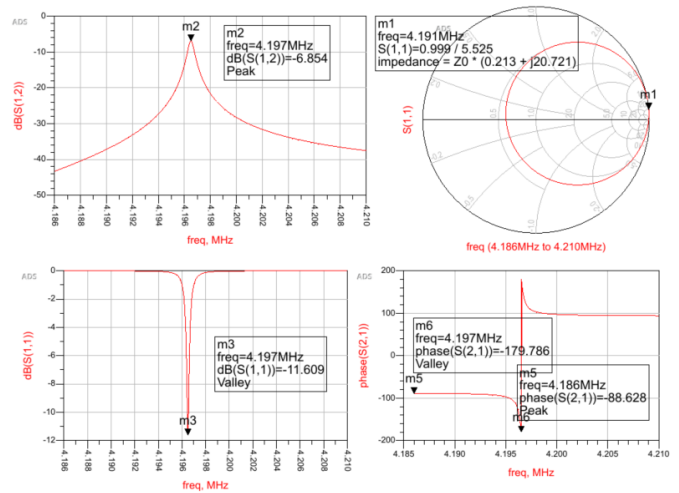

Fig. 7 ADS simulation plots -with frequency detuned to 4.197MHz from 4.196MHz

## RF frequency mixer for IF generation from RF and LO

In order to generate IF signals from RF a level-7 RF mixer (Minicircuits ZEM-4300) is used with an LO signal generated from an RF source like a signal generator. The crystal filter circuit center frequency determines the IF frequency and LO= RF+IF. Two RF couplers are placed in the circuit between RF mixer and crystal filter board circuit to simulate signals from dual directional coupler at the input of super conducting RF cavity as shown in the schematic below.

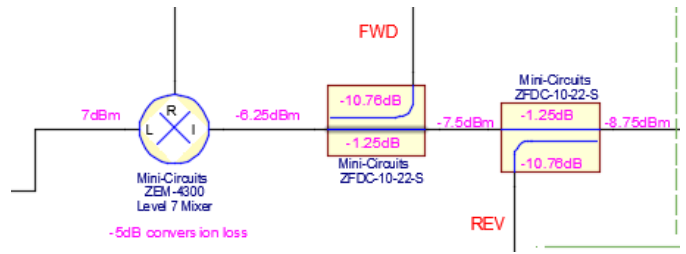

Fig. 8 Circuit drawing showing RF mixer and two couplers connected in series.

Frequency of 654.1943MHz is fed into the Local Oscillator at 10 dBm with a mixer to change the frequency of a signal to 650MHz. The center frequency of 650MHz signal input is down converted to an IF 4.1938470MHz. This frequency is passed through a low-pass crystal filter and is upconverted back to RF 650MHz using IQ modulation.

### I and Q Modulation for Upconversion of IF signal

In order to ensure that the final output of circuit is not dominated by LO leakage and other unwanted harmonics we use I and Q Modulator (EVAL-ADL5385) circuit instead of an RF frequency mixer. Output coming out of each signal chain forward, reverse and cavity is passed through a 90 ° hybrid splitter to generate in-phase and quadrature- phase components. These two IF I and Q components are modulated with LO to generate RF at 650MHz. A tunable bias circuit was designed to vary the bias voltage going to IBBN and QBBN inputs of the ADL5385. During the tuning process each potentiometer is adjusted going back and forth with IBBN and QBBN to achieve best LO leakage suppression in output. from IF to 650MHz. At 650 MHz, calculated Q= 650 MHz/50 Hz = 13 x $10^6$. In order to obtain RF Signal from IF up conversion signal we made a design choice of using an IQ modulator instead of RF mixer is used to avoid having to use narrow band high frequency filters to suppress LO leakage. EVAL-ADL5385 (30 MHz to 2200 MHz Quadrature Modulator) as shown in Fig. 3 is used for I Q modulation to upconvert IF at 4.19MHz to RF at 650MHz and DC bias circuit compensates to suppress IF + RF= 4.19MHz + 650MHz = 654.19MHz. Adjusting the phase of Q compensates/suppress RF + 2 IF.

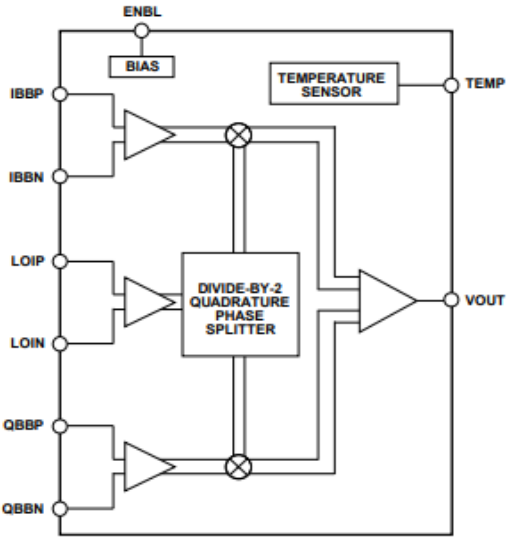

Fig.9 Block diagram of ADL5385 IQ Modulator

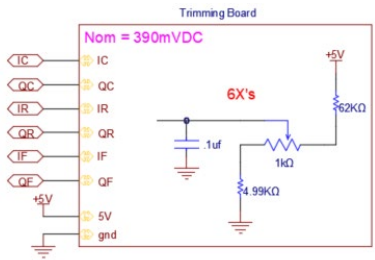

Fig.10 Circuit in the figure above is used to provide adjustable DC bias to I and Q inputs of the IQ modulator ADL5385.

### IV. ASSEMBLY

The cavity emulator circuit is assembled in a chassis, so the circuit is enclosed from external elements. Ports are provided in the front and back of the chassis to be able to connect easily with LLRF upconverters and downconverters. There is also a monitor port provided for monitoring LLRF RF drive signal. ADL5385 IQ modulator requires 5V DC for operation and this is provided externally from a power supply with a burndy pin connector.

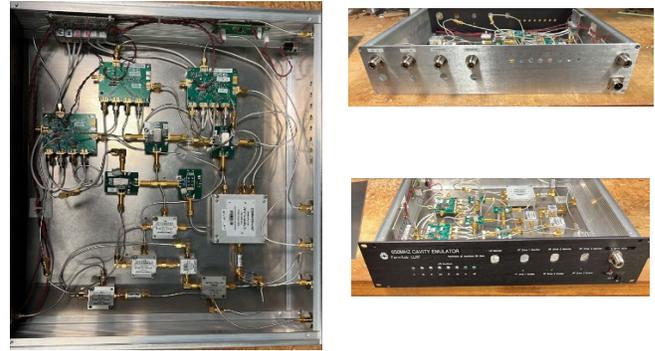

Fig.11 Pictures of top view, front view and rear view of the cavity emulator chassis.

### V. BENCH TEST RESULTS

Cavity emulator circuit is tested on a bench to validate it's performance benchmarking to a real super conducting cavity. Firstly, the crystal board which is the IF section of the circuit is measured to determine it's frequency and phase response. Bandwidth of the IF section would be similar to the overall bandwidth of the circuit.

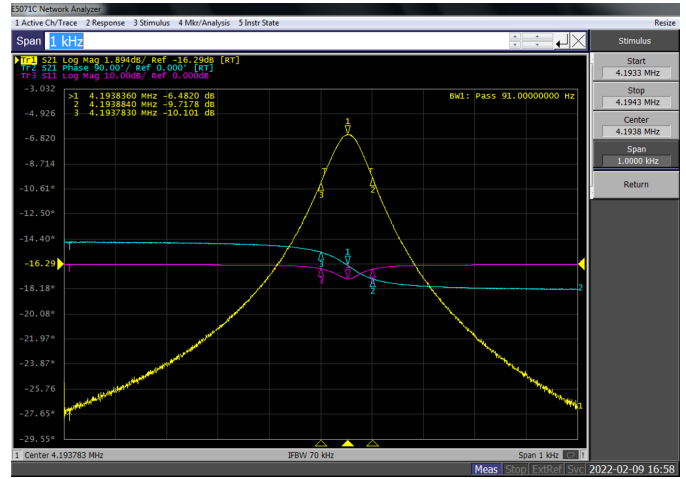

Fig.12 S-parameter measurement of cavity emulator crystal circuit

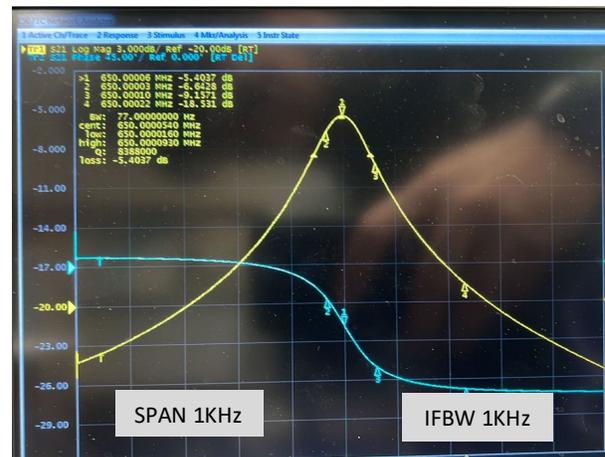

Fig.13 S-parameter measurement of entire cavity emulator cavity power channel.

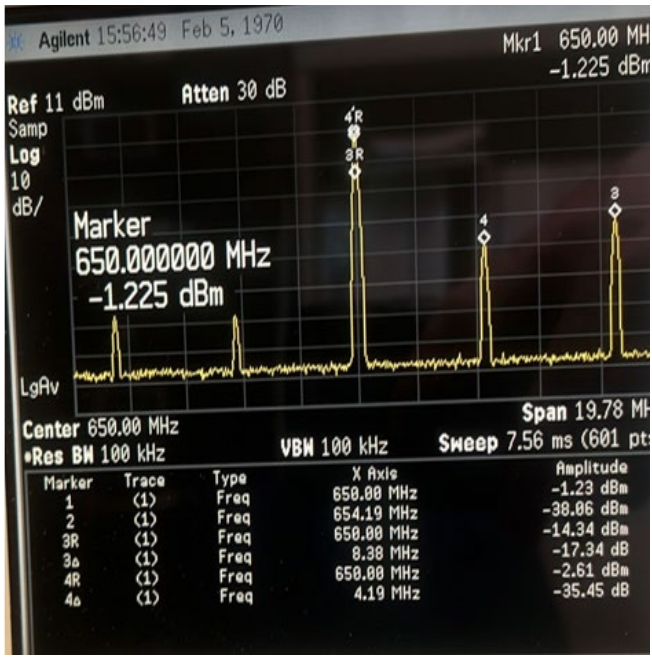

Fig.14 Frequency spectrum of cavity emulator -cavity power channel

## VI. TEST STAND APPLICATION

The cavity emulator chassis is used in the check-out of LLRF system at PIP-II test stands at Fermilab. We

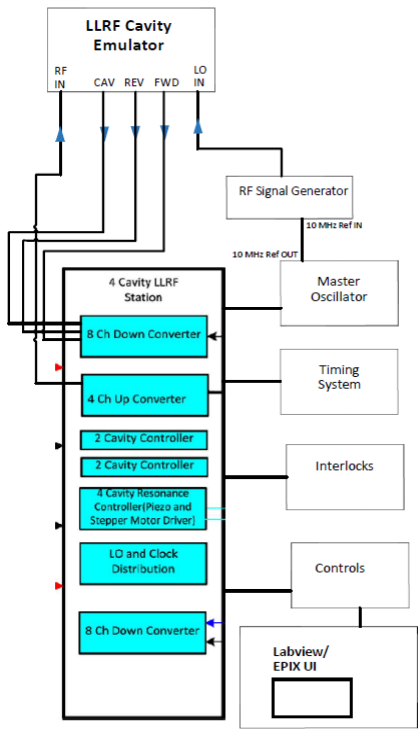

Fig.15 Frequency spectrum of cavity emulator -cavity power channel

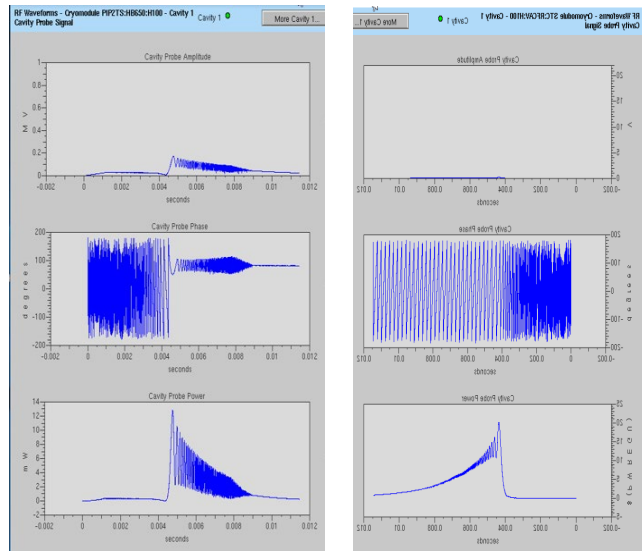

Fig.16 (L) Cavity emulator response, (R) superconducting cavity response

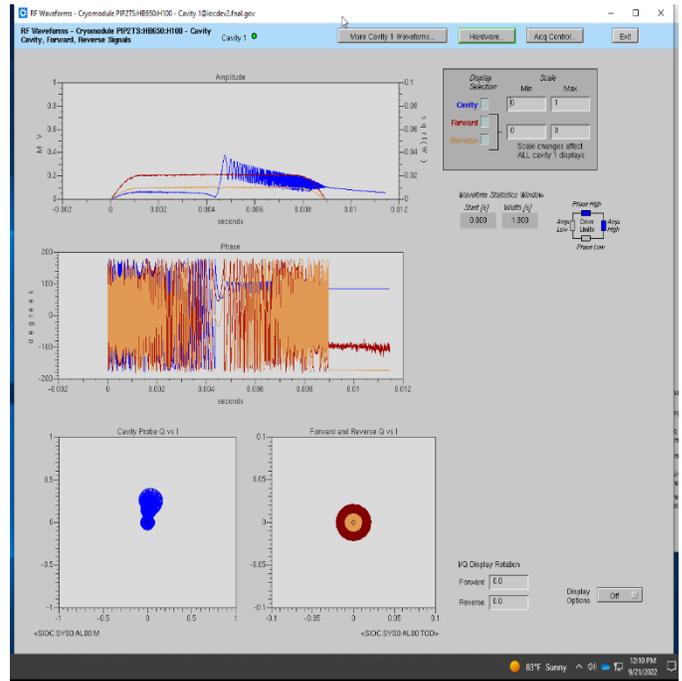

Fig.17 Response of a cavity emulator for a chirp signal

## CONCLUSION

The analog cavity emulator system described in this study is comparatively inexpensive, small, simple to produce, adaptable, wideband, and frequency independent. By altering the values of the standard components, the frequency and bandwidth can be changed to that of a required super conducting cavity to be emulated and desired response obtained.